\documentclass[doublecol]{epl2} 

\usepackage{amsmath}
\usepackage{amssymb}

\usepackage{graphicx}
\graphicspath{{./}{./images/}{./images_all/}{./images_all/single_phase_dynamics/}}

\title{Dynamical systems study in single-phase multiferroic materials}

\author{Kuntal Roy\thanks{E-mail: \email{royk@purdue.edu}}}
\shortauthor{Kuntal Roy}

\institute{                    
  School of Applied and Engineering Physics, Cornell University, Ithaca, New York 14853, USA
	\thanks{Present Address: School of Electrical and Computer Engineering, Purdue University, West Lafayette, Indiana 47907, USA}
}


\abstract{
Electric field induced magnetization switching in single-phase multiferroic materials is intriguing for both fundamental studies and potential technological applications. Here we develop a framework to study the switching dynamics of coupled polarization and magnetization in such multiferroic materials. With the coupling term between the polarization and magnetization as an invariant dictated by the Dzyaloshinsky-Moriya vector, the dynamical systems study reveals switching failures and oscillatory mode of magnetization if the polarization and magnetization relax slowly during switching. 
}

\begin{document}

\maketitle

\section{Introduction} \label{sec:introduction} Multiferroics usually represent materials that are both ferroelectric and ferromagnetic~\cite{RefWorks:664,RefWorks:164,RefWorks:853,RefWorks:854,RefWorks:558,RefWorks:521,RefWorks:322,RefWorks:855,RefWorks:856,RefWorks:711,RefWorks:859,RefWorks:858}. Such materials in single-phase were usually thought to be rare~\cite{RefWorks:512}, and hence multiferroic composites in 2-phase, i.e., a ferroelectric layer strain-coupled to a ferromagnet, are usually deemed to be the replacement~\cite{RefWorks:558,roy13_spin,roy13,roy14,roy14_2,roy14_4,roy11,roy13_2}. However, there have been recent resurgence of interests~\cite{RefWorks:665,RefWorks:707} and some mechanisms of coupling polarization and magnetization in single-phase materials are coming along~\cite{RefWorks:695,RefWorks:669,RefWorks:667}. This can lead to possible technological applications~\cite{RefWorks:435} of switching a bit of information (stored in the magnetization direction) by an electric field~\cite{RefWorks:709}. This eliminates the need to switch magnetization by a cumbersome magnetic field or spin-polarized current~\cite{roy13_spin}, although new concepts are being investigated e.g., utilizing giant spin-Hall effect~\cite{roy14_3}. The electric field switches the polarization and the intrinsic coupling between the polarization and magnetization switches the magnetization between its 180$^\circ$ symmetry equivalent states. One way to couple polarization and magnetization that has taken attention is due to Dzyaloshinsky-Moriya (DM) interaction~\cite{RefWorks:697,RefWorks:698}, which arises due to spin-orbit correction to Anderson's superexchange~\cite{RefWorks:745}. This is called ferroelectrically induced weak Ferromagnetism (wFM), in which two magnetic sublattices of an antiferromagnet cant in a way to produce a residual magnetization~\cite{RefWorks:699,RefWorks:700,RefWorks:701,RefWorks:695,RefWorks:702}.

\begin{figure*}
\centering
\includegraphics[width=0.8\textwidth]{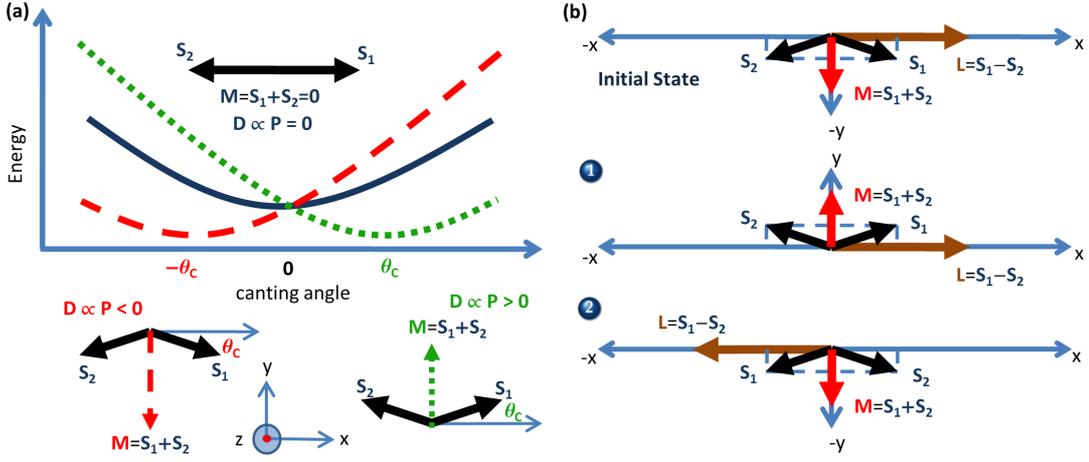}
\caption{(a) Energy of the two spins in a predominantly antiferromagnetic configuration with respect to canting angle $\theta_c$. The canting happens due to Dzyaloshinsky-Moriya (DM) interaction. The DM vector is proportional to polarization, i.e., $\mathbf{D} \propto \mathbf{P}$ and acts in the $z$-direction. When $\theta_c$ is positive, the net magnetization \textbf{M} points up ($+y$-axis), and if $\theta_c$ is negative, the net magnetization \textbf{M} points down ($-y$-axis). Without any canting, the net magnetization is zero as in the case of a perfect antiferromagnet. (b) Magnetization initially points along the $-y$-xis.  To respect the invariant dictated by the DM interaction [$\propto \mathbf{P}\cdot(\mathbf{L}\times\mathbf{M})$], with the reversal of polarization $\mathbf{P}$, either the magnetization \textbf{M} (case 1) or the AFM vector \textbf{L}  (case 2) may flip.}
\label{fig:schematic_AFM_M}
\end{figure*}

While first-principles calculations and experiments have been underway on the search of strongly-coupled multiferroic magnetoelectric materials possibly working at room-temperature, little have been studied on the dynamical nature of switching. The study of switching dynamics of magnetization in multiferroic composites, i.e., a piezoelectric layer strain-coupled to a magnetostrictive nanomagnet, have been very successful to understand the performance metrics, e.g., switching delay, energy dissipation, and switching failures~\cite{roy13_spin,roy13,roy14_2}. Here, the switching dynamics of polarization is studied by forming a Hamiltonian system with Landau-Ginzburg functional~\cite{RefWorks:723}, while the magnetization dynamics is studied by the Landau-Lifshitz-Gilbert (LLG) equation of motion~\cite{RefWorks:162,RefWorks:161}. We focus on magnetization switching due to electric field induced polarization switching, i.e., converse magnetoelectric (ME) effect for technological applications rather than the switching dynamics due to direct ME effect. We particularly consider the dynamics in single-domains with an eye to achieve high-density of devices rather than considering domain walls in higher dimensions, for which we need to consider the competition between the exchange interaction and dipole coupling among the spins~\cite{RefWorks:158}. {Note that switching dynamics in BiFeO$_3$ has been studied using a first-principles-based effective Hamiltonian within molecular dynamics simulations~\cite{RefWorks:852,RefWorks:861}.} Here we perform a \emph{comprehensive} analysis in emerging strongly-coupled multiferroics {(ferroelectrically-induced wFM \emph{by design} dictated by DM interaction~\cite{RefWorks:695})} by varying different parameters in the LLG dynamics. The analysis of switching dynamics reveals very significant motion of magnetization when polarization is switched by an electric field. It is shown that magnetization may fail to switch or even can go into an oscillatory state of motion. The phenomenological damping parameter for both polarization and magnetization plays a crucial role in shaping the dynamics of the coupled polarization-magnetization in these multiferroic materials.

\section{Model} 

We consider two spins, one representative to each magnetic sublattice of an antiferromagnet, to build up the present model. The dynamics of the two spins $\mathbf{S_1}$ and $\mathbf{S_2}$ can be described by the Landau-Lifshitz-Gilbert (LLG) equation~\cite{RefWorks:162,RefWorks:161} as follows:
\begin{align}
\frac{d\mathbf{S_1}}{dt} &= - |\gamma'|\,\mathbf{S_1} \times \mathbf{H_{S1}} - \frac{\alpha |\gamma'|}{S}\, \mathbf{S_1} \times \left( \mathbf{S_1} \times \mathbf{H_{S1}} \right) 
\label{eq:LLG_S1}\\
\frac{d\mathbf{S_2}}{dt} &= - |\gamma'|\,\mathbf{S_2} \times \mathbf{H_{S2}} - \frac{\alpha |\gamma'|}{S}\, \mathbf{S_2} \times \left( \mathbf{S_2} \times \mathbf{H_{S2}} \right),
\label{eq:LLG_S2}
\end{align}
\noindent
where $\mathbf{H_{S1}}$ and $\mathbf{H_{S2}}$ are the effective fields on the spins $\mathbf{S_1}$ and $\mathbf{S_2}$, respectively, defined as $\mathbf{H_{S1}} = - ({\partial \mathcal{H}/\partial \mathbf{S_1}})$ and $\mathbf{H_{S2}} = - (\partial \mathcal{H}/\partial \mathbf{S_2})$, $\mathcal{H}$ is the potential energy of the two spin system, expressed as 
\begin{equation}
\mathcal{H}=-J\, \mathbf{S_1}\cdot\mathbf{S_2} - \mathbf{D}\cdot (\mathbf{S_1} \times \mathbf{S_2}) - K S_{1,z}^2 - K S_{2,z}^2,
\label{eq:energy_term}
\end{equation}
\noindent
$J$ denotes the exchange coupling between the spins, $\mathbf{D}$ is the Dzyaloshinsky-Moriya (DM) vector~\cite{RefWorks:697,RefWorks:698} (here, we will consider the case when the vector $\mathbf{D}$ points along perpendicular to the plane ($x$-$y$ plane) on which the spins reside, i.e., along the $z$-direction and proportional to polarization $\mathbf{P}$, which is also in the $z$-direction~\cite{RefWorks:695}) expressed as $\mathbf{D} = D(t)\,\mathbf{\hat{e}_z}$ making 
\begin{equation}
D(t) \propto P(t),
\label{eq:D_prop_P}
\end{equation}
$K$ is the single-ion anisotropy constant, $\gamma'=\gamma/(1+\alpha^2)$, $\gamma$ is the gyromagnetic ratio of electrons, $\alpha$ is the phenomenological Gilbert damping constant~\cite{RefWorks:161}, and $S = |\mathbf{S_1}| = |\mathbf{S_2}|$. As required, it is possible to include the long-range interaction too in the energy term~\cite{RefWorks:802}. The net magnetization $\mathbf{M}$ and the antiferromagnetic (AFM) vector $\mathbf{L}$ for the two spin system are $\mathbf{M} = \mathbf{S_1} + \mathbf{S_2}$ and $\mathbf{L} = \mathbf{S_1} - \mathbf{S_2}$, respectively. Note that the following two identities hold: $\mathbf{M}\cdot\mathbf{L} = 0$ and $\mathbf{M}^2 + \mathbf{L}^2 = 4S^2$.

The polarization dynamics is based on the Landau-Ginzburg functional~\cite{RefWorks:723}
\begin{equation}
G = \left[-\frac{a_1}{2}\,P^2 + \frac{a_2}{4}\,P^4\right] - \mathbf{E}.\mathbf{P}, 
\label{eq:polarization_functional}
\end{equation}
where $a_1$ and $a_2$ are the ferroelectric coefficients (both are greater than zero) and $\mathbf{E}=E\,\mathbf{\hat{e}_z}$ is the applied electric field that switches the polarization $\mathbf{P}$ in the $z$-direction. We assume single-domain case~\cite{RefWorks:781} and follow the prescription in Ref.~\cite{RefWorks:723} to trace the trajectory of polarization. Note that polarization is switched by moving ions, which couples to the magnetization dynamics via the DM term $\mathbf{D}$ [see Eq.~\eqref{eq:energy_term}]. On the other hand, rotation of spins does not quite move the heavy ions affecting the polarization.

\begin{figure*}
\centering
\includegraphics[width=0.8\textwidth]{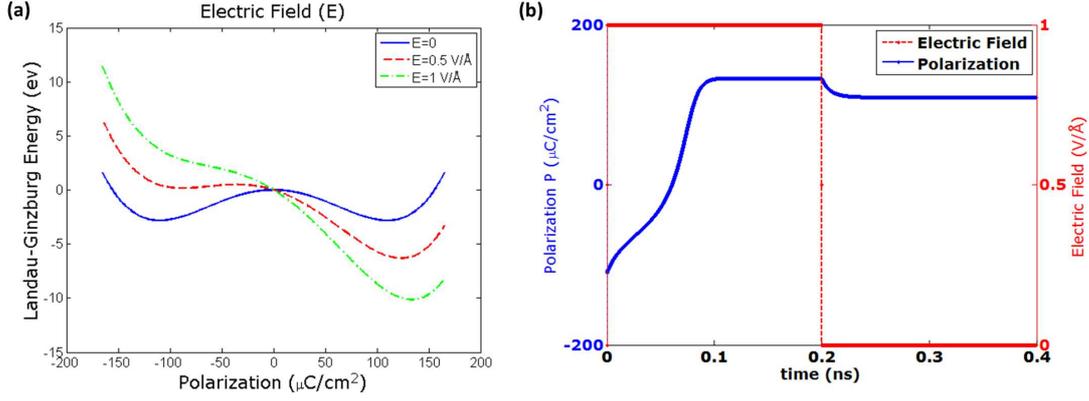}
\caption{(a) The potential landscape of polarization with electric field as a parameter [Equation~\eqref{eq:polarization_functional}]. Note that it requires a critical electric field to topple the barrier between polarization's two 180$^\circ$ symmetry equivalent states. (b) Switching of polarization with the application of an electric field. Initially, the polarization was pointing towards $-z$-axis. With the application of electric field, the polarization does not reach instantly towards the $+z$-axis, how fast the polarization relaxes to the minimum energy position depends on the polarization damping. Note that after the withdrawal of electric field, the polarization direction is maintained, i.e., the switching is non-volatile. The slight increase of polarization over $P_s$ is due to the application of electric field [see the potential landscapes in part (a)], which can be followed from the Equation~\eqref{eq:polarization_functional} too.}
\label{fig:dynamics_polarization_potential}
\end{figure*}

Figure~\ref{fig:schematic_AFM_M}a depicts how the \emph{asymmetric} Dzyaloshinsky-Moriya (DM) interaction can lead to two anti-parallel magnetization directions, i.e., $180^\circ$ symmetry equivalent states. Depending on the sign of the canting angle $\theta_c$ of the spins, the direction of the DM vector changes and the energy expression as in the Equation~\eqref{eq:energy_term} gives rise to two magnetization states in opposite directions. The DM vector $\mathbf{D}$ is proportional to polarization $\mathbf{P}$ and hence, if we switch the polarization, two cases can happen to respect the invariant due to DM interaction $\mathbf{P}\cdot(\mathbf{L}\times\mathbf{M})$ [$\propto \, \mathbf{D} \cdot (\mathbf{S_1} \times \mathbf{S_2})$  term in Equation~\eqref{eq:energy_term}]: (1) The magnetization $\mathbf{M}$ can change the direction (i.e., switches successfully), and (2) The AFM vector $\mathbf{L}$ may change the direction (i.e., $\mathbf{M}$ fails to switch). The two cases are depicted in the Fig.~\ref{fig:schematic_AFM_M}b.

\section{Results and Discussions} 

We consider a perovskite system $NiTiO_3$~\cite{RefWorks:695,RefWorks:720} in R3c space group~\cite{RefWorks:747} as a prototype to analyze the switching dynamics. Although, $NiTiO_3$ in R3c space group is not yet experimentally realized, the concept of polarization-magnetization coupling predicted from group theory is promising. The parameters are chosen as follows: saturation polarization $P_s = 110$ $\mu$C/cm$^2$, ferroelectric coefficients $a_1=1.568\times 10^{10}$ V\,m/C, $a_2=1.296\times 10^{10}$ V\,m/C, polarization damping $\beta = 0.286$ V\,m\,Sec/C (that switches the polarization in realistic time 100 ps~\cite{RefWorks:430}, see Fig.~\ref{fig:dynamics_polarization_potential}), $S = 1.6 \, \mu_B$, $M = 0.25\, \mu_B$, $J=-2.2$ meV, $D_s=0.35$ meV [corresponding to $P_s$, i.e., $D(t)=(D_s/P_s)\,P(t)$], $K = -0.03$ meV~\cite{RefWorks:695}. We will consider that the electric field switches the polarization from $-P_s$ to $+P_s$ in the $z$-direction.

The magnetization damping, through which magnetization relaxes to the minimum energy position, can have a wide range of values (10$^{-4}$ -- 0.8) by modifying the spin-orbit strength, doping etc. and it can be determined by ferromagnetic resonance (FMR), magneto-optical Kerr effect, x-ray absorption spectroscopy, and spin-current driven rotation with the addition of a spin-torque term~\cite{RefWorks:650,RefWorks:834,RefWorks:812}. Hence, we focus on investigating the magnetization dynamics for a wide range of phenomenological damping parameter.

\begin{figure}
\centering
\includegraphics{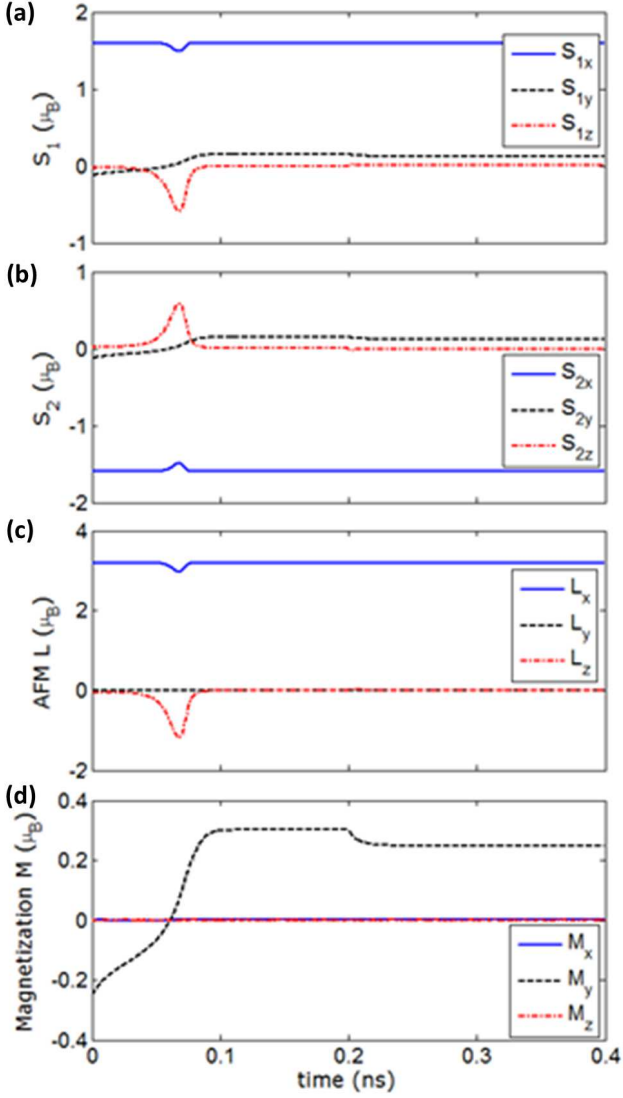}
\caption{Dynamics of magnetization for damping parameter $\alpha=0.1$. Magnetization does switch successfully. (a) Dynamics of $\mathbf{S_1}$, (b) Dynamics of $\mathbf{S_2}$, (c) Dynamics of AFM vector $\mathbf{L}$, and (d) Dynamics of magnetization $\mathbf{M}$.}
\label{fig:dynamics_alpha_0d1}
\end{figure}

We will initially assume the single-ion anisotropy $K=0$ and we will see later the consequence of considering it. Figure~\ref{fig:dynamics_alpha_0d1} shows the dynamics of magnetization when damping parameter is on the higher side, e.g., 0.1. We see that magnetization has switched successfully in the end (see Fig.~\ref{fig:dynamics_alpha_0d1}d), while the AFM vector did not switch (see Fig.~\ref{fig:dynamics_alpha_0d1}c). Note that the spins $\mathbf{S_1}$ and $\mathbf{S_2}$ are deflected from the $x$-$y$ plane, in the $z$-direction due to rotational motion of magnetization. Also, note that magnetization's $x$- and $z$-component and AFM vector's $y$-component have not changed at all due to the complimentary dynamics of the spins $\mathbf{S_1}$ and $\mathbf{S_2}$. This corresponds to the case (1) in Fig.~\ref{fig:schematic_AFM_M}b.

\begin{figure}
\centering
\includegraphics{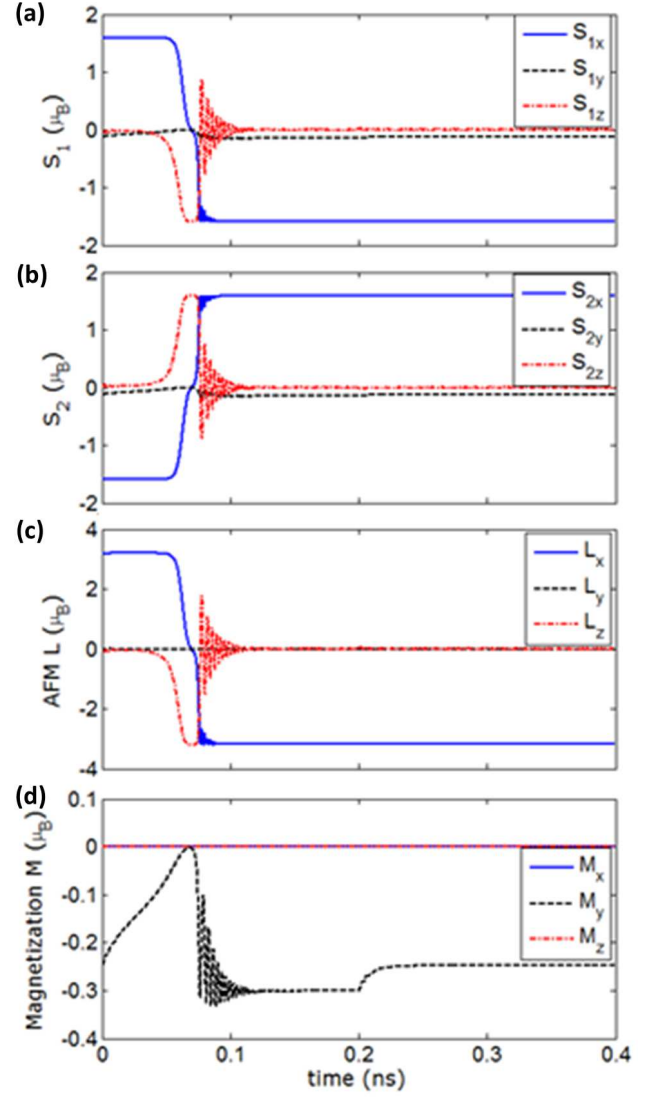}
\caption{Dynamics of magnetization for damping parameter $\alpha=0.01$. Magnetization fails to switch successfully. (a) Dynamics of $\mathbf{S_1}$, (b) Dynamics of $\mathbf{S_2}$, (c) Dynamics of AFM vector $\mathbf{L}$, and (d) Dynamics of magnetization $\mathbf{M}$.}
\label{fig:dynamics_alpha_0d01}
\end{figure}

Figure~\ref{fig:dynamics_alpha_0d01} plots the dynamics when magnetization damping $\alpha=0.01$. We see that magnetization has failed to switch (see Fig.~\ref{fig:dynamics_alpha_0d01}d), while the AFM vector is switched successfully (see Fig.~\ref{fig:dynamics_alpha_0d01}c). Magnetization was on the way to change its direction, but eventually, magnetization came back to its initial state. This corresponds to the case (2) in Fig.~\ref{fig:schematic_AFM_M}b. Due to low damping, we notice ringing in all the plots in the Fig.~\ref{fig:dynamics_alpha_0d01}.

For the lower damping of $\alpha=0.01$, from the simulation results as shown in the Fig.~\ref{fig:dynamics_alpha_0d01}, the positions of the two spins have just got interchanged, which is depicted as the case (2) in Fig.~\ref{fig:schematic_AFM_M}b. Since the canting angle of the spins are small ($\theta_c \simeq 5^\circ$), one can say that the spins have rotated much more than that of the case for the higher damping of $\alpha=0.1$ [case (1) in Fig.~\ref{fig:schematic_AFM_M}b and the simulation results as shown in the Fig.~\ref{fig:dynamics_alpha_0d1}]. While both the cases as shown in the Fig.~\ref{fig:schematic_AFM_M}b respect the DM invariant at \emph{steady-state}, the \emph{dynamics} of magnetization dictates the final state that is reached. With a lower damping, the spins get deflected out-of-plane more (see the $z$-components of the spins $\mathbf{S_1}$ and $\mathbf{S_2}$ in the Figs.~\ref{fig:dynamics_alpha_0d1} and~\ref{fig:dynamics_alpha_0d01}) and this out-of-plane excursion eventually can lead the spins to interchange their positions as can be noticed from the Fig.~\ref{fig:dynamics_alpha_0d01}. The interchange of the spins $\mathbf{S_1}$ and $\mathbf{S_2}$ indeed respects the DM invariant [case (2) of Fig.~\ref{fig:schematic_AFM_M}b], but the magnetization $\mathbf{M}$ \emph{fails} to switch in this case, while the AFM vector $\mathbf{L}$ gets switched.

\begin{figure*}
\centering
\includegraphics[width=0.8\textwidth]{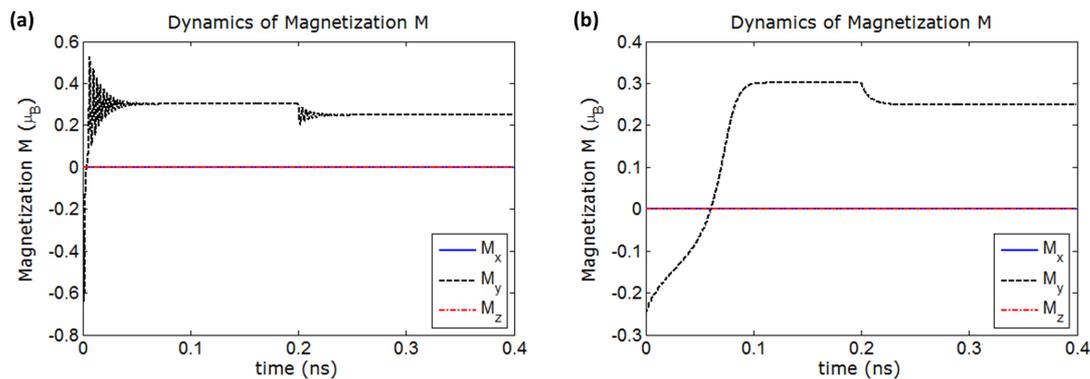}
\caption{Magnetization switches successfully at damping parameter $\alpha=0.01$. (a) Polarization is switched 200 times faster by changing the polarization damping. The magnetization switches successfully. (b) The single-ion anisotropy constant $K$ is taken into account. In this case, magnetization switches successfully too.}
\label{fig:dynamics_alpha_0d01_success}
\end{figure*}

To understand the magnetization dynamics further that how switching may be successful even at magnetization damping $\alpha=0.01$, we first investigate its dependence on the polarization dynamics. Simulation results show that magnetization switches successfully if we make the polarization damping 200 times faster [see Fig.~\ref{fig:dynamics_alpha_0d01_success}a]. Basically, due to the coupling between the polarization and magnetization, if polarization is switched faster, magnetization is also switched faster, which makes the switching successful. We further investigate the effect of single-ion anisotropy parameter $K$ on magnetization dynamics. The ion-anisotropy basically adds an extra field that tries to keep the magnetization in-plane (i.e., $x$-$y$ plane) and the simulation results show that magnetization switches successfully if we take the single-ion-anisotropy into account and the ringing in the magnetization dynamics does not show up in this case [see Fig.~\ref{fig:dynamics_alpha_0d01_success}b]. However, if the single-ion-anisotropy is reduced to a value of  $K = -0.003$ $\mu$eV, it is noticed that the magnetization fails to switch successfully.

We further study the effect of varying the DM interaction strength $D_s$ on the switching dynamics. It is found quite obviously that as $D_s$ decreases for a fixed canting angle $\theta_c$, i.e., polarization-magnetization coupling weakens, the AFM vector $\mathbf{L}$ deflects more, which can be interpreted as that the magnetization $\mathbf{M}$ is more prone to switching failures. For $\alpha=0.1$, if $D_s$ is reduced 10 times, the magnetization $\mathbf{M}$ still switches successfully.

\begin{figure*}[t]
\centering
\includegraphics[width=0.8\textwidth]{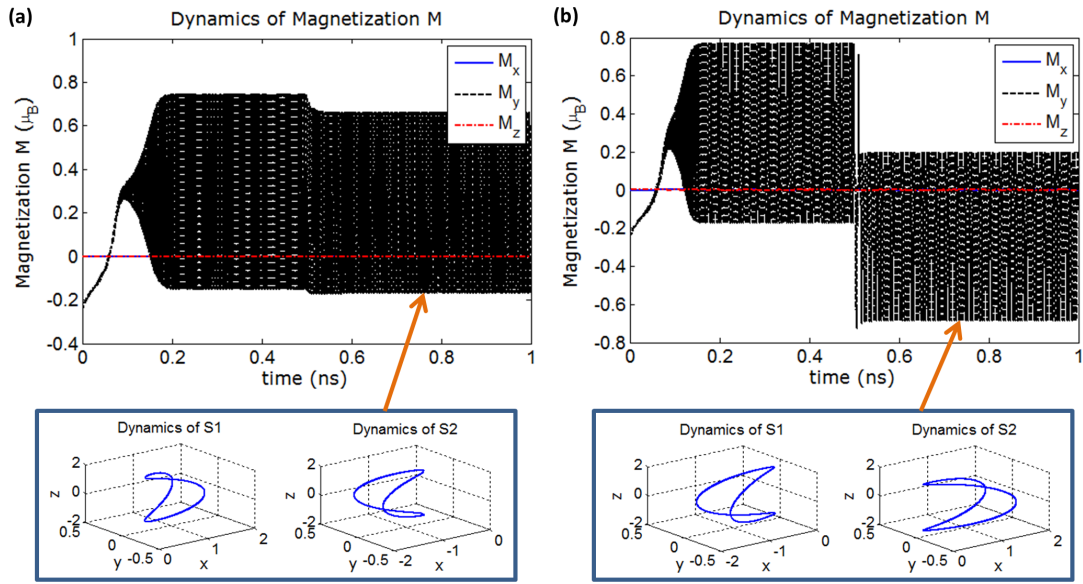}
\caption{Magnetization oscillation for low damping parameters. The electric field is withdrawn at 0.5 ns. (a) Damping parameter $\alpha=0.001$. Magnetization has got past from $-y$-axis to $+y$-axis but could not settle there rather oscillates around that position with a time period of 3.3 ps. (b) Damping parameter $\alpha=10^{-4}$. Magnetization got past from $-y$-direction to $+y$-direction but could not settle there rather oscillates around that position with a time period of 5.3 ps. With the removal of electric field, the magnetization traverses toward $-y$-direction and oscillates around there. The three-dimensional dynamics of the spins $\mathbf{S_1}$ and $\mathbf{S_2}$ are shown at the lower part of the figure for both the cases (a) and (b).}
\label{fig:dynamics_oscillation_alpha_1e-3_1e-4}
\end{figure*}

An interesting investigation would be to see whether magnetization, being a rotational body, oscillates for a certain range of damping parameter. For example, a spin-polarized current can spawn oscillatory states in a nanomagnet~\cite{RefWorks:737}. Figure~\ref{fig:dynamics_oscillation_alpha_1e-3_1e-4}a shows the magnetization dynamics when damping parameter $\alpha=0.001$. The single-ion-anisotropy is included here and magnetization was able to get past towards the $+y$-direction, however, could not settle there and oscillates with a time period of 3.3 ps. With further lowering of the damping parameter, magnetization still oscillates, however with a lower frequency [see Fig.~\ref{fig:dynamics_oscillation_alpha_1e-3_1e-4}b].

The oscillatory mode of magnetization too can be explained from the out-of-plane excursion of the spins due to low damping. The spins get deflected out-of-plane (i.e., $z$-direction) and when they go completely out-of-plane, they continue rotating and reach the out-of-plane in the opposite directions than the previous ones. Therefore the spins sustain a \emph{self-oscillation}. The DM interaction ensures that the $x$- and $z$-component of the spins are canceled out and the $y$-components are added due to symmetry, as can be noticed in the Fig.~\ref{fig:dynamics_oscillation_alpha_1e-3_1e-4}. Note that such \emph{self-oscillation} occurs and sustains even in the absence of an external electric field, making the system \emph{unstable}. Such spontaneous \emph{self-oscillation} is not uncommon in electronic systems having negative damping due to positive feedback leading to instabilities. The oscillation time-period increases at a lower damping (see Fig.~\ref{fig:dynamics_oscillation_alpha_1e-3_1e-4}b) since it takes more time for magnetization to traverse for a lower damping parameter.

The research on single-phase multiferroic materials is still emerging, and the search for a room-temperature system that requires a low enough electric field for switching the polarization is still underway. It will be interesting to incorporate the thermal fluctuations in the model to understand the consequence on magnetization dynamics~\cite{RefWorks:186,roy13_spin}. Also for a shape-anisotropic single-domain nanomagnet, the corresponding anisotropy needs to be included for detailed simulation~\cite{RefWorks:158}.

\section{Conclusions} 

We have investigated the electric field induced magnetization switching dynamics in single-phase multiferroic materials. The dynamical system analysis, contrary to steady-state analysis, revealed important intriguing phenomena of switching failures and oscillatory mode of magnetization. The key parameters that can shape the dynamics of magnetization are identified. The phenomenological magnetization damping turns out to be a key parameter that can prevent successful switching of magnetization. Hence, the present analysis puts forward an important step toward analyzing magnetization switching dynamics between 180$^\circ$ symmetry equivalent states in the emerging multiferroic materials. Moreover, the analysis identifies the oscillatory mode of magnetization that can act as a source of microwave signals. The present analysis can facilitate designing multiferroic materials for relevant technological purposes.

\end{document}